\documentclass[reprint,amsmath,amssymb,aps,prd]{revtex4-2}
\usepackage[colorlinks,citecolor=purple,linkcolor=blue,anchorcolor=blue,filecolor=blue,
urlcolor=blue]{hyperref}
\usepackage{multirow}
\usepackage{tabularx}
\usepackage{graphicx}
\usepackage{ulem}
\usepackage{dcolumn}
\usepackage{bm}
\usepackage{todonotes}

\newcommand{\sym}{\mathrm{sym}}

\begin{document}

\title{Dark energy effects on realistic neutron stars}

\author{Juan M. Z. Pretel}
\email{juanzarate@cbpf.br}
\affiliation{Centro Brasileiro de Pesquisas F{\'i}sicas, Rua Dr.~Xavier Sigaud, 150 URCA, Rio de Janeiro CEP 22290-180, RJ, Brazil
}

\author{Sergio B. Duarte}
\email{sbd@cbpf.br}
\affiliation{Centro Brasileiro de Pesquisas F{\'i}sicas, Rua Dr.~Xavier Sigaud, 150 URCA, Rio de Janeiro CEP 22290-180, RJ, Brazil
}

\author{Jos\'e D. V. Arba\~nil}
\email{jose.arbanil@upn.pe}
\affiliation{Departamento de Ciencias, Universidad Privada del Norte, Avenida el Sol 461 San Juan de Lurigancho, 15434 Lima, Peru}
\affiliation{Facultad de Ciencias F\'isicas, Universidad Nacional Mayor de San Marcos, Avenida Venezuela s/n Cercado de Lima, 15081 Lima, Peru}

\author{Mariana Dutra}
\email{marianad@ita.br}
\affiliation{Departamento de F\'isica e Laborat\'orio de Computa\c c\~ao Cient\'ifica Avan\c cada e Modelamento (Lab-CCAM), Instituto Tecnol\'ogico de Aeron\'autica, DCTA, 12228-900, S\~ao Jos\'e dos Campos, SP, Brazil
}

\author{Odilon Lourenço}
\email{odilon.ita@gmail.com}
\affiliation{Departamento de F\'isica e Laborat\'orio de Computa\c c\~ao Cient\'ifica Avan\c cada e Modelamento (Lab-CCAM), Instituto Tecnol\'ogico de Aeron\'autica, DCTA, 12228-900, S\~ao Jos\'e dos Campos, SP, Brazil
}

\date{\today}

\begin{abstract}
By considering realistic equations of state (EoSs) to describe the ordinary matter of the stellar crust, in this study, we explore the effect of a dark energy core, made of Chaplygin Dark Fluid (CDF), on neutron stars (NSs). To accomplish this purpose, we solve the stellar structure equations and investigate the impact of the CDF parameters on the several macroscopic properties of NSs such as mass-radius ($M-R$) relation, and tidal deformabilities of a single star and of a binary system, the latter being of great importance when analyzing gravitational-wave signals coming from the merger of such compact objects. We also present an analysis of the radial oscillation modes for the rapid phase transition, with the aim of distinguishing regions consisting of dynamically stable stars from those of unstable ones. Specifically, our outcomes reveal that an increase in the energy density jump (controlled by a parameter $\alpha$) leads to an increase in the radial stability of the NS with a CDF core. Furthermore, our theoretical results are consistent with the observational $M-R$ measurements of millisecond pulsars from NICER data and tidal deformability constraints from the GW170817 event.
\end{abstract}

\maketitle

\section{Introduction} 

According to various cosmological observations, our universe is currently undergoing an accelerated expansion phase \cite{Riess1998, Perlmutter1999, Jimenez2002, Eisenstein2005, Ade2016, Haridasu2017, Bengaly2020}, so that the late-time acceleration is a widely accepted phenomenon nowadays. The standard model of cosmology, also called the ``$\Lambda$CDM model'', is based on cold dark matter (responsible for cosmic structure formation) and cosmological constant $\Lambda$ which is commonly associated with the vacuum density energy of the universe. Such a model is in good agreement with the observational data \cite{Aghanim2020}, however, it suffers from coincidence and fine-tuning problems \cite{Weinberg1989, Padmanabhan2003}. Since then, many attempts have been proposed to address these issues and the literature therefore offers a variety of models as candidates to describe dark energy \cite{Ratra1988, Armendariz1999, Kamenshchik2001, Copeland2006, Li2011, Clifton2012, Joyce2016, Koyama2016, Montefalcone2020, Shankaranarayanan2022}.

A dark energy fluid with negative pressure and positive energy density may be a candidate to describe cosmic acceleration \cite{Kamenshchik2001}, although the exact nature of this dark energy still remains a mystery. In particular, negative pressure can be obtained using quintessence dark energy, where quintessence is more precisely a scalar field~\cite{Ratra1988}. Nevertheless, quintessence model fails to avoid fine-tuning in explaining the cosmic coincidence problem. In that regard, the Chaplygin fluid prescription provides a good alternative to explain the transition from a universe filled with dustlike matter in the early era to an accelerated expansion phase in the late stage \cite{Zheng2022}. In fact, the Chaplygin gas and its generalized models have been investigated and confronted with different observational cosmological measurements by several researchers \cite{Bento2003, Bertolami2004, Barreiro2008, Park2010, Xu2012, Wang2013, Li2019, Yang2019, Mamon2022}.

If it is assumed that the universe is filled with an exotic form of fluid, described by a Chaplygin-type equation of state (EoS), then dark energy could be present in stellar interiors. In recent years there has therefore been a growing interest in investigating the role of dark energy on the relativistic (internal and/or external) structure of compact objects. As a matter of fact, Anti-de Sitter black holes surrounded by Chaplygin Dark Fluid (CDF) were recently investigated \cite{Li2023, Sekhmani2024, SekhmaniPDU2024}, where the existence of a dark fluid described by the Chaplygin gas as a cosmic background is assumed. Additionally, Li and collaborators \cite{LiJCAP2024} explored the impact of CDF on the geodesic structure, shadows and optical appearance of a black hole surrounded by various profiles of accretions. In particular, the authors showed that the CDF parameter $B$ plays a crucial role in the formation of optical images of the black hole, by affecting features such as the event horizon radius, photon sphere radius, and the radius of the Innermost Stable Circular Orbit (ISCO)/OSCO.

In addition to black holes, the connection between dark energy and compact stars has become an active area of research in the last few years. Indeed, it was shown that Chaplygin-type EoS gives rise to compact-star radii and masses compatible with astrophysical observations \cite{Pretel2023EPJC, Tudeshki2023, Bhattacharjee2024, Jyothilakshmi2024}, and such stars are often referred to as dark energy stars (DESs) \cite{Rahaman2010, Panotopoulos2020EPJP, Panotopoulos2021, Tudeshki2023PLB}. These single-phase models have recently been extended to a hybrid context \cite{Pretel2024}, where dark energy is confined in the core of the compact star while the crust contains baryonic matter. The presence of the dark energy core (made of a CDF) leads to a sizable change in the internal structure of NSs with a polytropic EoS for the external layer. Despite the crust being described by a simple EoS, there is a great possibility that these stars exist in the universe since they are dynamically stable under radial perturbations and because their mass-radius relations are compatible with observational measurements. In this work, we aim to extend this simple toy model of hybrid stars by adopting more realistic EoSs to describe the crust and to investigate the influence of the dark energy core on their tidal deformability which are measurable in the chirp signals of merging NSs. To depict the ordinary matter of the outer layer, we use two well-known models, i.e., the relativistic mean-field (RMF) and the Skyrme one.

The arrangement of this paper is the following: in Sec.~\ref{Sec2} we provide both EoSs to depict our hybrid star model, one for the dark energy core and one for the crust made of hadronic matter. We employ parametrizations of two different hadronic models to describe the outer layer so that our stellar model is characterized by four astrophysical parameters $\{\rho_c, \rho_{\rm dis}^+, \alpha, A\}$. We introduce the basic differential equations related to spherically symmetric equilibrium configurations in the theory of General Relativity in Sec.~\ref{Sec3}. We continue our discussion of numerical calculations especially focusing on the mass-radius diagram, tidal deformability, and the radial stability for the rapid phase transition in Sec.~\ref{Sec4}. In the same section, we compare our theoretical predictions with observational data. Finally, we give a summary and our conclusions in Sec.~\ref{conclusion}.

\section{Equations of state}
\label{Sec2}

In a variety of astrophysical simulations, the EoS is a crucial input to solve the stellar structure equations. For a barotropic fluid, the EoS is nothing more than the functional relation between the energy density~$\rho$ and pressure~$p$, namely $p= p(\rho)$. To model our two-phase compact star, we follow the notation used in Ref.~\cite{Pretel2024} so that the EoS describing the entire star is given by
\begin{equation}\label{EoSEq}
    p(\rho) = 
  \begin{cases}
    A\rho_{\mbox{\tiny DE}}- \dfrac{B}{\rho_{\mbox{\tiny DE}}},  & \quad  0 \leq r \leq R_{\rm dis} ,  \\ 
    p_{\mbox{\tiny nm}}(\rho{\mbox{\tiny nm}}),  & \quad  R_{\rm dis} \leq r \leq R ,
  \end{cases}
\end{equation}
where $R_{\rm dis}$ is the radius of the discontinuous surface. In this expression, the indexes DE and NM refer to the quantities related to dark energy and normal matter, respectively. 

In order to describe the normal matter sector, we use two well-known models, namely, the relativistic mean-field~(RMF) and the Skyrme one. The former is described by the following Lagrangian density~\cite{baoanli08,dutra14,lattimer24}
\begin{align}
\mathcal{L} &= \overline{\psi}(i\gamma^\mu\partial_\mu - M_{\mbox{\tiny nuc}})\psi
+ g_\sigma\sigma\overline{\psi}\psi
- g_\omega\overline{\psi}\gamma^\mu\omega_\mu\psi 
\nonumber\\
&- \frac{g_\rho}{2}\overline{\psi}\gamma^\mu\vec{b}_\mu\vec{\tau}\psi
+ \frac{1}{2}(\partial^\mu \sigma \partial_\mu \sigma
- m^2_\sigma\sigma^2) - \frac{\mathcal{A}}{3}\sigma^3 - \frac{\mathcal{B}}{4}\sigma^4
\nonumber\\
& + \frac{1}{2}m^2_\omega\omega_\mu\omega^\mu + \frac{C}{4}(g_\omega^2\omega_\mu\omega^\mu){^2}
- \frac{1}{4}\vec{B}^{\mu\nu}\vec{B}_{\mu\nu} + \frac{1}{2}m^2_\rho\vec{b}_\mu\vec{b}^\mu
\nonumber\\
&+ g_\sigma g_\rho^2\sigma\vec{b}_\mu\vec{b}^\mu
\left(\alpha_2+\frac{1}{2}{\alpha'_2}g_\sigma\sigma\right)
+ \frac{1}{2}{\alpha'_3}g_\omega^2 g_\rho^2\omega_\mu\omega^\mu
\vec{b}_\mu\vec{b}^\mu
\nonumber\\
&+ g_\sigma g_\omega^2\sigma\omega_\mu\omega^\mu
\left(\alpha_1+\frac{1}{2}{\alpha'_1}g_\sigma\sigma\right) -\frac{1}{4}F^{\mu\nu}F_{\mu\nu},
\label{rmf}
\end{align}
in which $\psi$ is the Dirac spinor representing the nucleon field; $\sigma$, $\omega^\mu$, and $\vec{b}_\mu$ describe the scalar, vector, and isovector fields corresponding to the respective mesons $\sigma$, $\omega$, and~$\rho$. The antisymmetric tensors are $F_{\mu\nu}=\partial_\nu\omega_\mu-\partial_\mu\omega_\nu$ and
$\vec{B}_{\mu\nu}=\partial_\nu\vec{b}_\mu-\partial_\mu\vec{b}_\nu$. The nucleon and mesons masses are $M_{\mbox{\tiny nuc}}$, $m_\sigma$, $m_\omega$, and $m_\rho$. Finally, the free parameters of the model are $g_\sigma$, $g_\omega$, $g_\rho$, $\mathcal{A}$, $\mathcal{B}$, $C$, $\alpha_1$, $\alpha_1'$, $\alpha_2$, $\alpha_2'$ and~$\alpha_3'$. The Hartree approximation to the $T_{00}$ and $T_{ii}/3$ components of the energy-momentum tensor leads to the energy density and pressure written as (units of $\hbar=c=1$)
\begin{align}
&\rho_{\mbox{\tiny RMF}} = \frac{1}{2}m^2_\sigma\sigma^2 
+ \frac{\mathcal{A}}{3}\sigma^3 + \frac{\mathcal{B}}{4}\sigma^4 - \frac{1}{2}m^2_\omega\omega_0^2 
- \frac{C}{4}(g_\omega^2\omega_0^2)^2 \nonumber\\
&- \frac{1}{2}m^2_\rho b_{0(3)}^2
+g_\omega\omega_0n+\frac{g_\rho}{2} b_{0(3)}n_3 + \rho_{\mbox{\tiny kin}}^p + \rho_{\mbox{\tiny kin}}^n
\nonumber\\
&- g_\sigma g_\rho^2\sigma b_{0(3)}^2 
\left(\alpha_2+\frac{1}{2}{\alpha'_2} g_\sigma\sigma\right) - \frac{1}{2}{\alpha'_3}g_\omega^2 g_\rho^2\omega_0^2 b_{0(3)}^2
\nonumber \\
&- g_\sigma g_\omega^2\sigma\omega_0^2\left(\alpha_1+\frac{1}{2}{\alpha'_1}g_\sigma\sigma\right),
\label{dermf}
\end{align}
and
\begin{align}
&p_{\mbox{\tiny RMF}} = - \frac{1}{2}m^2_\sigma\sigma^2 - \frac{\mathcal{A}}{3}\sigma^3 -
\frac{\mathcal{B}}{4}\sigma^4 + \frac{1}{2}m^2_\omega\omega_0^2 
\nonumber\\
&+ \frac{C}{4}(g_\omega^2\omega_0^2)^2 + \frac{1}{2}m^2_\rho b_{0(3)}^2 
+ p_{\mbox{\tiny kin}}^p + p_{\mbox{\tiny kin}}^n
\nonumber\\
&+ g_\sigma g_\rho^2\sigma b_{0(3)}^2 
\left(\alpha_2+\frac{1}{2}{\alpha'_2} g_\sigma\sigma\right)
+ \frac{1}{2}{\alpha'_3}g_\omega^2g_\rho^2\omega_0^2 b_{0(3)}^2
\nonumber\\
&+ g_\sigma g_\omega^2\sigma\omega_0^2\left(\alpha_1+\frac{1}{2}{\alpha'_1}g_\sigma\sigma\right),
\label{presrmf}
\end{align}
with kinetic terms reading
\begin{align}
\rho_{\mbox{\tiny kin}}^{p,n}&=\frac{1}{\pi^2}\int_0^{{k_F}_{p,n}}
\hspace{-0.2cm}k^2(k^2+M^{*2})^{1/2}dk,
\\
p_{\mbox{\tiny kin}}^{p,n} &=
\frac{1}{3\pi^2}\int_0^{{k_F}_{p,n}}
\hspace{-0.5cm}\frac{k^4dk}{(k^2+M^{*2})^{1/2}},
\end{align}
in which $n=n_p+n_n$, $n_3=n_p-n_n$, ${k_F}_{p,n}$ is the Fermi momentum of the nucleon, and $M^*$ is the nucleon effective mass. From $\rho_{\mbox{\tiny RMF}}$, one obtains the chemical potentials of protons and neutrons. Such expressions are given by
\begin{align}
\mu_{p,n}^{\mbox{\tiny RMF}} &= (k_{Fp,n}^2+{M^*}^2)^{1/2} + g_\omega\omega_0 \pm \frac{g_\rho}{2} b_{0(3)},
\end{align}
with $+$~($-$) for protons~(neutrons). In the above expressions, $\sigma$, $\omega_0$ and $b_{0(3)}$ are the ``classical'' mean-field versions of the respective fields $\sigma$, $\omega^\mu$, and $\vec{b}_\mu$. $n_p$~($n_n$)~is the proton~(neutron) density.

Another widely used model nowadays in the description of hadronic matter is the zero-range
nonrelativistic interaction known as Skyrme model~\cite{stone07,dutra12,lattimer24}, that  treats many-nucleon systems through effective density-dependent $NN$ and $NNN$ interactions instead of realistic ones. In this context, details of the two and three-body forces are not explicitly taken into account, since all relevant physical information is considered in a phenomenological way by the free parameters of the model. Once again, the use of the mean-field Hartree-Fock approximation leads to
\begin{align}
\rho_{\mbox{\tiny sky}} &= \frac{3}{10M_{\mbox{\tiny nuc}}}\left(\frac{3\pi^2}{2}\right)^{2/3}n^{5/3}H_{5/3} \nonumber\\
&+ \frac{t_0}{8}n^2[2(x_0+2)-(2x_0+1)H_2] \nonumber \\
&+ \frac{1}{48}\sum_{i=1}^{2}t_{3i}n^{\sigma_{i}+2} [2(x_{3i}+2)-(2x_{3i}+1)H_2]
\nonumber\\
&+ \frac{3}{40}\left(\frac{3\pi^2}{2}\right)^{2/3}n^{8/3}\left(aH_{5/3}+bH_{8/3}\right) + nM_{\mbox{\tiny nuc}}
\label{desky}
\end{align}
and
\begin{align}
p_{\mbox{\tiny sky}}&= \frac{1}{5M_{\mbox{\tiny nuc}}}\left(\frac{3\pi^2}{2}\right)^{2/3}n^{5/3}H_{5/3} \nonumber\\
&+\frac{t_0}{8}n^2[2(x_0+2)-(2x_0+1)H_2]\nonumber\\
&+\frac{1}{48}\sum_{i=1}^{3}t_{3i}(\sigma_i+1)n^{\sigma_i+2}
[2(x_{3i}+2)-(2x_{3i}+1)H_2]\nonumber\\
&+\frac{1}{8}\left(\frac{3\pi^2}{2}\right)^{2/3}n^{8/3}\left(aH_{5/3}+bH_{8/3}\right),
\label{pressky}
\end{align}
with
\begin{align}
a&=t_1(x_1+2)+t_2(x_2+2),\\
b&=\frac{1}{2}\left[t_2(2x_2+1)-t_1(2x_1+1)\right],
\label{eq:b} \\
H_l(y)&=2^{l-1}[y^l+(1-y)^l],
\end{align}
where $y=n_p/n$ is the proton fraction of the system. The set of free paramters in this case is $x_0$, $x_1$, $x_2$, $x_{31}$, $x_{32}$, $x_{33}$, $t_0$, $t_1$, $t_2$, $t_{31}$, $t_{32}$, and $t_{33}$. The nucleon chemical potential is given by
\begin{align}
&\mu_{p,n}^{\mbox{\tiny sky}} = \frac{1}{2M_{\mbox{\tiny nuc}}}\left(\frac{3\pi^2}{2}\right)^{2/3}n^{2/3}H_{5/3}(y)
\nonumber\\
&+ \frac{1}{5}\left(\frac{3\pi^2}{2}\right)^{2/3}n^{5/3}[aH_{5/3}(y) + bH_{8/3}(y)]
\nonumber\\
&+\frac{t_0}{4}n[2(x_0+2)-(2x_0+1)H_2(y)]
\nonumber\\
&+ 
\frac{1}{48}\sum_{i=1}^{3}t_{3i}(\sigma_i+2)n^{\sigma_i+1}[2(x_{3i}+2)-(2x_{3i}
+1)H_2(y)]
\nonumber\\
&\pm \frac{1}{2}\left[1 \mp 
(2y-1)\right]\left\{\frac{3}{10M_{\mbox{\tiny nuc}}}\left(\frac{3\pi^2}{2}\right)^{2/3} 
n^{2/3}H'_{5/3}(y) \right.
\nonumber\\
&\left.-\frac{t_0}{8}n(2x_0+1)H'_2(y) 
-\frac{1}{48}\sum_{i=1}^{3}t_{3i}n^{\sigma_i+1}(2x_{3i} +1)H'_2(y) \right.
\nonumber\\
&+\left. \frac{3}{40} 
\left(\frac{3\pi^2}{2}\right)^{2/3}n^{5/3}[aH'_{5/3}(y) + bH'_{8/3}(y)]\right\} + M_{\mbox{\tiny nuc}},
\label{muqsk}
\end{align}
where $H'_l(y)=dH_l/dy$. Upper (lower) sign for proton (neutron).

At this point, we need to use parametrizations of the presented hadronic models. In particular, we choose the BSR8~\cite{bsr8} and SLy4~\cite{sly4} ones. The respective nuclear matter properties, at the saturation density, are given in Table.~\ref{tab:nep}. 
\begin{table}[!htb]
\tabcolsep=0.04cm
\centering
\caption{Nuclear empirical properties, at the saturation density, related to the models used in this work, namely, saturation density ($n_0$), binding energy ($B_e$), incompressibility ($K_0$), ratio of the nucleon effective mass over the rest mass ($M^*_0/M_{\rm nuc}$), symmetry energy ($E_\sym$), symmetry energy slope ($L_\sym$), and symmetry energy curvature ($K_\sym$).}
\begin{tabular}{l c c c c c c c}
\hline
model & $n_0$       & $B_0$    & $K_0$ & $M^*_0/M_{\rm nuc}$ & $E_\sym$ & $L_\sym$ & $K_\sym$ \\
      & [fm$^{-3}$] & [MeV]    & [MeV] &                   & [MeV]    & [MeV]    & [MeV]      \\ 
\hline 
BSR8  & $0.147$     & $-16.04$ & $231$ & $0.61$            & $31.08$ & $60$      & $-1$       \\
Sly4  & $ 0.160$    & $-15.97$ & $230$ & $0.69$            & $32.00$ & $46$      & $-120$     \\
\hline
\end{tabular}
\label{tab:nep}
\end{table}
Such parametrizations were chosen due to their capability of reproducing charge radii, ground state binding energies, and giant monopole resonances of a set of the following spherical nuclei: $^{16}\rm O$, $^{34}\rm Si$, $^{40}\rm Ca$, $^{48}\rm Ca$, $^{52}\rm Ca$, $^{54}\rm Ca$, $^{48}\rm Ni$, $^{56}\rm Ni$, $^{78}\rm Ni$, $^{90}\rm Zr$, $^{100}\rm Sn$, $^{132}\rm Sn$, and $^{208}\rm Pb$. In addition, they are also in agreement with the macroscopic properties of neutron stars. We address the reader to Ref.~\cite{brett-jerome} for a systematic study performed with 415 relativistic mean field and nonrelativistic Skyrme-type interactions in this regard. 

In order to generate cold compact stars with hadronic core, one needs to impose $\beta$ equilibrium and charge neutrality. These conditions are satisfied since leptons are included in the system. More specifically, here we refer to electrons~($e$) and muons~($\mu$). Therefore, the equations $\mu_n-\mu_p=\mu_e=\mu_\mu$ and $n_p=n_e+n_\mu$ are established, and the respective solutions are used to define the final equations of state, total energy density, and total pressure, given by 
\begin{align}
\rho_{\mbox{\tiny total}} = \rho_{\mbox{\tiny hm}} + \frac{\mu_e^4}{4\pi^2} 
+ \frac{1}{\pi^2}\int_0^{\sqrt{\mu_\mu^2-m^2_\mu}}\hspace{-0.6cm}dk\,k^2(k^2+m_\mu^2)^{1/2},
\label{eq:totaled}
\end{align}
and
\begin{align} 
p_{\mbox{\tiny total}} = p_{\mbox{\tiny hm}} + \frac{\mu_e^4}{12\pi^2} +\frac{1}{3\pi^2}\int_0^{\sqrt{\mu_\mu^2-m^2_\mu}}\hspace{-0.5cm}\frac{dk\,k^4}{(k^2+m_\mu^2)^{1/2}},
\label{eq:totalp}
\end{align}
where $\rho_{\mbox{\tiny hm}}$ and $p_{\mbox{\tiny hm}}$ stand for the hadronic contribution to the energy density and pressure. Such terms are described by the Skyrme model ($\rho_{\mbox{\tiny hm}}=\rho_{\mbox{\tiny sky}}$, $p_{\mbox{\tiny hm}}=p_{\mbox{\tiny sky}}$, $\mu_{p,n}=\mu^{\mbox{\tiny sky}}_{p,n}$) or the relativistic mean field one ($\rho_{\mbox{\tiny hm}}=\rho_{\mbox{\tiny RMF}}$, $p_{\mbox{\tiny hm}}=p_{\mbox{\tiny RMF}}$, $\mu_{p,n}=\mu^{\mbox{\tiny RMF}}_{p,n}$), both of them aforementioned. Moreover, the electron density~$n_e$ is related to~$\mu_e$ through $n_e=\mu_e^3/(3\pi^2)$, and the muon density is~$n_\mu=[(\mu_\mu^2 - m_\mu^2)^{3/2}]/(3\pi^2)$ for $\mu_\mu\geq m_\mu=105.7$~MeV, and $n_\mu=0$ otherwise. 

For a complete description of the hadronic part of the compact star, we include in the normal matter sector a suitable equation of state that describes the outer crust, namely, the one proposed by Baym, Pethick, and Sutherland (BPS)~\cite{bps}. We adopt this prescription in a density region of $6.3\times10^{-12}\,\mbox{fm}^{-3}\leqslant n\leqslant2.5\times10^{-4}\,\mbox{fm}^ {-3}$. Therefore, the final expressions for energy density and pressure read
\begin{align}
\rho_{\mbox{\tiny nm}} = \rho_{\mbox{\tiny total}} + \rho_{\mbox{\tiny BPS}},
\label{eq:ednm}
\end{align}
and
\begin{align} 
p_{\mbox{\tiny nm}} = p_{\mbox{\tiny total}} + p_{\mbox{\tiny BPS}},
\label{eq:pnm}
\end{align}
respectively.

On the other hand, the core of the compact star is described by a Chaplygin-type EoS, with $A$ and $B$ being positive constants. Notice that the negative term ``$-B/\rho_{\mbox{\tiny DE}}$'' represents the original Chaplygin gas model and leads to cosmic acceleration \cite{Kamenshchik2001}. In addition to the various cosmological motivations, this gas also has interesting characteristics from the point of view of $d$-brane theory. Indeed, it was shown that the original CDF can be obtained from the string Nambu–Goto action for $d$-branes moving in a $(d+2)$-dimensional spacetime \cite{Bordemann1993, Ogawa2000, Jackiw2000}.

At the phase-splitting surface, the pressure must be continuous so that Eq.~(\ref{EoSEq}) implies
\begin{equation}\label{EqB}
    B = A(\rho_{\rm dis}^+)^2 - (\rho_{\rm dis}^+)p_{\mbox{\tiny nm}}(\rho_{\rm dis}^-) ,
\end{equation}
where the energy density jump allows us to define $\alpha = \rho_{\rm dis}^-/\rho_{\rm dis}^+ \leq 1$ as the ratio of energy densities at the radial coordinate separating the two phases \cite{Sotani2001, Pretel2024}. In the literature, $\alpha$ is known as the density jump parameter \cite{Arbanil2023}, and the energy density is obviously continuous when $\alpha=1$. The smaller $\alpha$, the larger the energy density jump for a fixed value of $\rho_{\rm dis}^+$. 

\begin{figure}[!htb]
\centering
\includegraphics[scale=0.72]{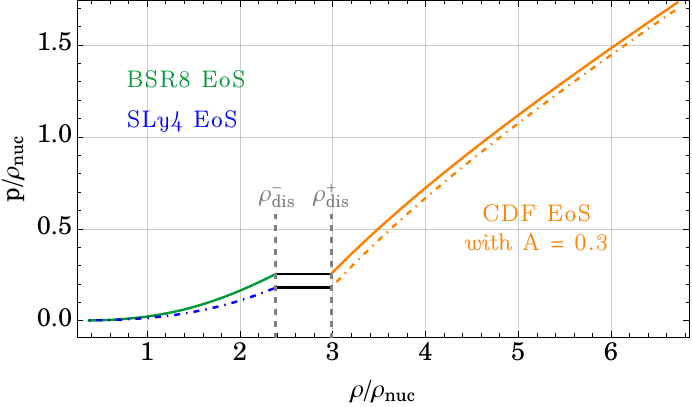}
\caption{Equation of state given by Eq.~\eqref{EoSEq} for our hybrid star model by considering inner energy density $\rho_{\rm dis}^+= 0.8 \times 10^{15}\, \rm g/cm^3$, ratio of densities at the discontinuous surface $\alpha= 0.8$, CDF parameter $A= 0.3$. The maximum central energy density in this case is $\rho_c= 1.8 \times 10^{15}\rm g/cm^3$. Note that both, pressure and energy density, have been normalized by the energy density of normal symmetric nuclear matter $\rho_{\rm nuc}= 2.68\times 10^{14}\, \rm g/cm^3$ (found by using the values of $M_{\mbox{\tiny nuc}}=939$~MeV, $B_0=-16$~MeV and $n_0=0.163$~fm$^{-3}$ for the nucleon mass, binding energy, and saturation density, respectively).}
\label{FigEoSs}
\end{figure}

In summary, once the EoS for normal matter $p_{\mbox{\tiny nm}}= p_{\mbox{\tiny nm}}(\rho_{\mbox{\tiny nm}})$ is adopted and a central density value $\rho_c$ is given as in single-phase stars, our hybrid star model with a CDF core will be described under the choice of the three free parameters $\{\rho_{\rm dis}^+, \alpha, A\}$. The values adopted for these parameters will be based on recent studies \cite{Arbanil2023, Pretel2024}, in addition to allowing us to obtain results compatible with the different recent observational measurements. For some specific values of these parameters, Fig.~\ref{FigEoSs} shows the EoS with phase transition for our NS model with a CDF core.

\begin{figure*}[!htb]
\centering
\includegraphics[scale=0.70]{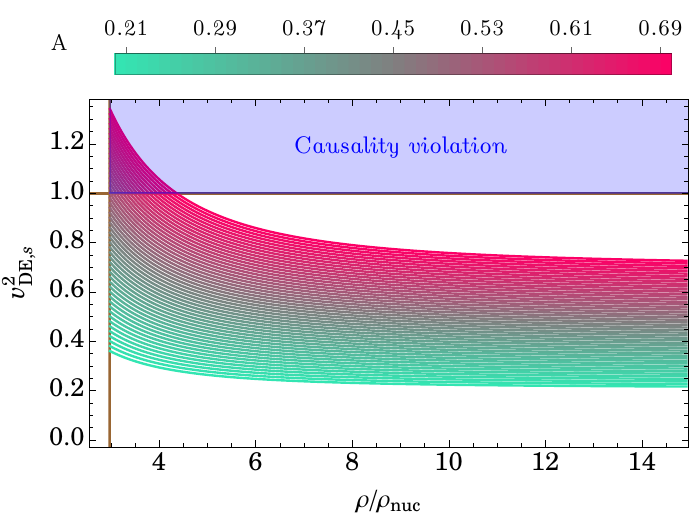}
\includegraphics[scale=0.70]{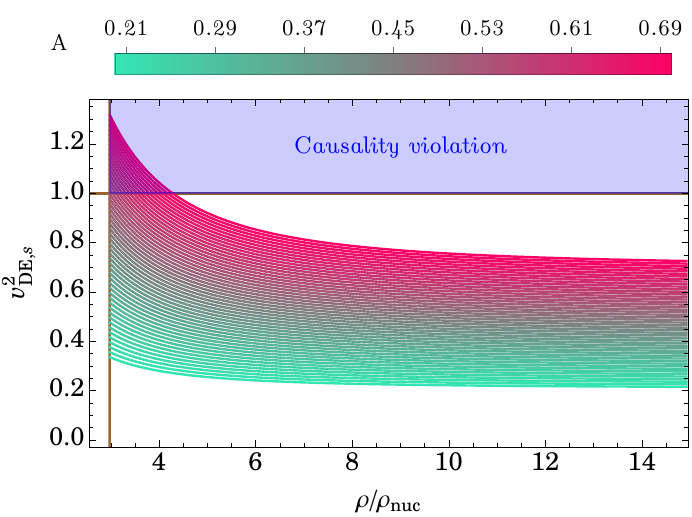}
\caption{Squared speed of sound for the dark energy fluid confined in the core of the star considering a crust whose microphysics is described by the SLy4 EoS (left panel) and BSR8 EoS (right panel). We also assumed $\rho_{\rm dis}^+= 0.8 \times 10^{15}\, \rm g/cm^3$, $\alpha= 0.8$ and CDF parameter $A \in [0.2, 0.7]$. The blue region indicates superluminal signal propagation, which imposes an upper limit for $A$ (see also Table \ref{tableUppeLimA} for other values of $\alpha$). }
\label{FigSpeedSound}
\end{figure*}

It is important to remark that the squared speed of sound, defined by $v_{\rm s}^2= dp/d\rho$, must be less than $1$ in order to respect the causality condition. In the case of single-phase dark energy stars, such condition implies that $A<0.5$ since $v_{\rm DE,s}^2(R)= 2A <1$ \cite{Pretel2023EPJC}. Nonetheless, in our current stellar model, the dark energy fluid is confined only to the core of the star and hence the maximum value of $A$ can be larger than 0.5. From Eq.~(\ref{EqB}), we see that this restriction depends on the other model parameters as well as the EoS describing the envelope. In view of Eq.~(\ref{EoSEq}) for the core of the star, we have
\begin{equation}
    v_{\rm DE,s}^2 = A + \frac{B}{\rho_{\rm DE}^2},
\end{equation}
which is shown in Fig.~\ref{FigSpeedSound} as a function of the energy density for $\alpha= 0.8$, $\rho_{\rm dis}^+= 0.8 \times 10^{15}\, \rm g/cm^3$ and a wide range of values for the CDF parameter $A \in [0.2,0.7]$. 
We have included the full range of energy densities that generate the $M-R$ diagrams in Figs.~\ref{FigMRdSLy4} and \ref{FigMRdDEBSR8}. It is clearly observed that some values of $A$ lead to a speed of sound higher than the speed of light, i.e., they allow superluminal signal propagation. For the causality condition to be obeyed, $A\lesssim 0.530$ for NSs with SLy4 EoS crust, while $A\lesssim 0.542$ for the envelope with BSR8 EoS. This indicates that a stiffer EoS for the crust allows slightly larger upper limits for $A$. Constraints on $A$ for other values of~$\alpha$ can be found in Table \ref{tableUppeLimA}. 
\begin{table}
\begin{minipage}[b]{70mm}
\caption{\label{tableUppeLimA} Upper limit for the CDF parameter $A$ fulfilling the causality condition $v_{\rm DE,s}^2 <1$, where we have used the two EoS models for the envelope of the hybrid star with $\rho_{\rm dis}^+= 0.8 \times 10^{15}\, \rm g/cm^3$. }
\begin{ruledtabular}
\begin{tabular}{|c|p{2.4cm}|p{2.4cm}|}
  $\alpha$  &  SLy4 EoS crust  &  BSR8 EoS crust  \\ 
\hline
  0.7  &  $A\lesssim 0.521$  &  $A\lesssim 0.530$  \\
  0.8  &  $A\lesssim 0.530$  &  $A\lesssim 0.542$  \\
  0.9  &  $A\lesssim 0.542$  &  $A\lesssim 0.556$  \\
  1.0  &  $A\lesssim 0.556$  &  $A\lesssim 0.571$  
\end{tabular}
\end{ruledtabular}
\end{minipage}
\end{table}
Furthermore, we have verified that the ordinary matter EoSs describing the crust does not allow superluminal signal propagation and, therefore, all EoSs adopted in this work satisfy the causality condition along all equilibrium configurations on the $M-\rho_c$ curves.

\section{Compact star properties}
\label{Sec3}

As usual, we assume that the stellar source is described by the energy-momentum tensor of a perfect fluid, i.e., $T_\mu^\nu = {\rm diag}(-\rho, p, p, p)$, where $\rho$ is the energy density and $p$ is the pressure. Furthermore, for a spherically symmetric system, the line element is given by
\begin{eqnarray}\label{metric}
  ds^2 = - e^{2\Phi}dt^2 + e^{2\Psi}dr^2 + r^2(d\theta^2 + \sin^2\theta d\phi^2 ),
\end{eqnarray}
where the functions $\Phi(r)$ and $\Psi(r)$ are metric variables to be determined. In view of the Einstein field equation, $R_{\mu\nu} - g_{\mu\nu}\mathcal{R}/2 = 8\pi T_{\mu\nu}$ (we choose units with $G = c = 1$), the mass-radius diagram is obtained by numerically integrating the Tolman-Oppenheimer-Volkoff (TOV) equations 
\begin{align}
  \frac{dm}{dr} &= 4\pi r^2 \rho , \label{TOV1}\\
  \frac{dp}{dr} &= - (\rho+ p)\, \frac{m+ 4\pi r^3 p }{r\left(r - 2m\right)}, \label{TOV2}
\end{align}
with $m(r)$ being the gravitational mass function inside a sphere of radius $r$, and is related to the metric function $\Psi(r)$ by means of $e^{-2\Psi} = 1- 2m/r$. The surface of the star is reached when the pressure vanishes $p(R)= 0$, so that $R$ denotes the radius of the fluid sphere and its total mass is $M= m(R)$.

Other important properties of compact stars such as tidal deformability and oscillation frequencies require determining the metric function $\Phi(r)$. It is obtained after solving the differential equation
\begin{equation}
    \frac{d\Phi}{dr}= -\frac{1}{\rho+ p}\frac{dp}{dr}  \label{DiffEqPhi} ,
\end{equation}
where $e^{2\Phi(R)} = 1- 2M/R$ has to be satisfied at the stellar surface due to the Schwarzschild exterior solution.

Neutron stars are tidally deformed under the presence of a companion star, and this deformation can be inferred through gravitational waves emitted from the late phase of the inspiral of a binary compact star system \cite{Most2018, Chatziioannou2020, Dietrich2021}. Therefore, here we also examine the tidal deformability for our NS models with dark energy in the core. The tidal perturbation $y(r)$ is governed by the following equation
\begin{equation}\label{yEq}
    r\frac{dy}{dr} = -y^2 + (1 - r\mathcal{P})y - r^2\mathcal{Q} ,
\end{equation}
with the initial condition $y(0)=2$ \cite{Postnikov2010}, and 
\begin{align}
    \mathcal{P} =&\ \frac{2}{r} + e^{2\Psi}\left[ \frac{2m}{r^2} + 4\pi r(p - \rho) \right] ,  \\
    \mathcal{Q} =&\ 4\pi e^{2\Psi}\left[ 5\rho + 9p + \frac{\rho+ p}{dp/d\rho} \right] - \frac{6e^{2\Psi}}{r^2} - 4\Phi'^2 .  \label{EqforQ}
\end{align}

There is a jump in the energy density across the phase-splitting surface, but the pressure is the same between these two different values of energy density. At $r= R_{\rm dis}$, we shall therefore notice that Eq.~(\ref{EqforQ}) exhibits a singularity due to the term $dp/d\rho$. In that regard, for first-order transitions in hybrid stars, it has been shown that the perturbation function $y(r)$ must satisfy the following junction condition at the interface \cite{Postnikov2010, Janos2020, Arbanil2023}
\begin{equation}\label{BCTidalDef}
    [y]_-^+ = \frac{4\pi R_{\rm dis}^3(\rho_{\rm dis}^- - \rho_{\rm dis}^+)}{m(R_{\rm dis}) + 4\pi R_{\rm dis}^3 p(R_{\rm dis})} , 
\end{equation}
where we have used the notation $[z]_-^+ = z^+ - z^-$, with $z$ standing for any variable that crosses the interface. Once the surface value $y_R= y(R)$ is calculated, we proceed to obtain the tidal Love number $k_2$ via expression
\begin{align}
k_2 &= \frac{8}{5}(1- 2C)^2C^5 \left[ 2C(y_R -1) - y_R+ 2 \right]  \nonumber  \\
&\times \left\lbrace 2C[ 4(y_R+ 1)C^4 + (6y_R- 4)C^3 \right.  \nonumber  \\
&\left.+\ (26- 22y_R)C^2 + 3(5y_R -8)C - 3y_R+ 6 \right]   \nonumber  \\
&\left.+\ 3(1-2C)^2\left[ 2C(y_R- 1)- y_R +2 \right]\ln(1-2C) \right\rbrace^{-1} ,
\label{LoveNumEq}
\end{align}
where $C= M/R$ is the compactness of the star of mass $M$ and radius $R$. Consequently, the dimensionless tidal deformability is determined from $\Lambda= 2k_2C^{-5}/3$.

A rigorous analysis of the radial stability of a compact star involves determining the vibration mode frequencies when the star is subjected to radial perturbations. These perturbations have a harmonic time dependence $\sim e^{i\omega t}$, where $\omega$ is the oscillation frequency to be calculated. 
For the stars generated from our approach to exist in nature, they have to be dynamically stable. The differential equations governing such radial pulsations were first derived by Chandrasekhar \cite{ChandrasekharApJ, ChandrasekharPRL} and for numerical convenience, they can be written in the form \cite{Gondek1997, Vasquez2010, Pretel2020MNRAS, Hong2023}
\begin{align}
  \frac{d\zeta}{dr} &= -\frac{1}{r}\left( 3\zeta + \frac{\Delta p}{\gamma p} \right) + \frac{d\Phi}{dr}\zeta ,  \label{ROEq1}   \\
  \frac{d(\Delta p)}{dr} &=\ \zeta\left\lbrace \omega^2e^{2(\Psi - \Phi)}(\rho + p)r - 4\frac{dp}{dr} \right.  \nonumber  \\
  &\left. - 8\pi (\rho + p)e^{2\Psi}rp + r(\rho + p)\left(\frac{d\Phi}{dr}\right)^2 \right\rbrace   \nonumber \\
  & - \Delta p \left[ \frac{d\Phi}{dr} + 4\pi(\rho + p)re^{2\Psi} \right] ,  \label{ROEq2} 
\end{align}
where the perturbation variable $\zeta$ is related to the Lagrangian displacement $\xi$ through $\zeta= \xi/r$. Furthermore, $\Delta p$ is the Lagrangian perturbation of the fluid pressure, and $\gamma= (1+\rho/p)dp/d\rho$ is the adiabatic index.

The boundary conditions at the center and surface of single-phase stars hold for hybrid stars, i.e., $\Delta p= -3\zeta\gamma p$ as $r\rightarrow 0$, and $\Delta p =0$ as $r\rightarrow R$. The former condition arises from ensuring regular solutions in any region of the star and the latter comes from the fact that the pressure vanishes at the surface. However, our NS model contains two separated phases and suitable junction conditions on the discontinuous surface have to be used when solving the pulsation equations. Such conditions, dependent on the velocity of the phase transition near the domain wall, were derived by Pereira \textit{et al.}~\cite{Pereira2018}. 

For slow phase transitions, the timescale of the process transforming one phase into another is much larger than those of the radial perturbations, and there is no mass transfer between the two phases. For this type of transition, the junction conditions at the discontinuous wall are as follows
    \begin{align}\label{JuncCond1}
        \left[ \zeta \right]_-^+ &= 0,  &  \left[ \Delta p \right]_-^+ &= 0. 
    \end{align}

Meanwhile, for rapid phase transitions, the characteristic timescale for transforming one phase into another is much smaller than that of the perturbations. This means that there is a mass transfer from one phase to the other at the phase-splitting boundary, involving the matching conditions
\begin{align}\label{JuncCond2}
        \left[ \zeta- \frac{\Delta p}{rdp/dr} \right]_-^+ &= 0,  &  \left[ \Delta p \right]_-^+ &= 0.
    \end{align}
In the next section, we only investigate the radial oscillation modes for the rapid phase transition case.

\section{Results}\label{Sec4}

\begin{table*}
\begin{minipage}[b]{160mm}
\caption{\label{table} Maximum-mass configurations for NSs with a CDF core using an inner energy density $\rho_{\rm dis}^+= 0.8 \times 10^{15}\, \rm g/cm^3$ and two realistic EoSs for the normal matter envelope. Note further that we have assumed three values for $A$ and four values for $\alpha$, as in Figs.~\ref{FigMRdSLy4} and \ref{FigMRdDEBSR8}. The critical values $\rho_c$ shown for $A= 0.4$ correspond precisely to the central densities where the squared vibration frequency $\omega_0^2$ vanishes in the upper plots of Fig.~\ref{FigFreqs}.}
\begin{ruledtabular}
\begin{tabular}{|c|c|cccc|cccc|}
\multicolumn{2}{|c|}{Parameters}  &  \multicolumn{4}{c|}{SLy4 EoS crust}  &  \multicolumn{4}{c|}{BSR8 EoS crust}  \\
\hline
$A$  &  $\alpha$  &  $\rho_c\, [10^{15}\, \rm g/cm^3]$  &  $R\, [\rm km]$  &  $M\, [M_\odot]$  &  $\Lambda$  &  $\rho_c\, [10^{15}\, \rm g/cm^3]$  &  $R\, [\rm km]$  &  $M\, [M_\odot]$  &  $\Lambda$  \\
\hline
  \multirow{4}{*}{$0.2$}  &  0.7  &  2.980  &  9.461  &  1.399  &  47.235  &  2.550  &  10.344  &  1.496  &  59.477  \\
  &  0.8  &  2.725  &  9.873  &  1.459  &  56.936  &  2.231  &  10.920  &  1.594  &  67.068  \\
  &  0.9  &  2.467  &  10.271  &  1.532  &  62.570  &  1.970  &  11.390  &  1.697  &  67.366  \\
  &  1.0  &  2.250  &  10.610  &  1.611  &  62.587  &  1.828  &  11.686  &  1.791  &  60.581  \\
\hline
  \multirow{4}{*}{$0.3$}  &  0.7  &  2.739  &  9.875  &  1.753  &  13.123  &  2.515  &  10.481  &  1.820  &  15.582  \\
  &  0.8  &  2.599  &  10.162  &  1.797  &  15.898  &  2.313  &  10.886  &  1.889  &  18.368  \\
  &  0.9  &  2.446  &  10.442  &  1.850  &  18.029  &  2.147  &  11.225  &  1.961  &  19.786  \\
  &  1.0  &  2.302  &  10.691  &  1.908  &  19.173  &  2.019  &  11.480  &  2.030  &  19.906  \\
\hline
  \multirow{4}{*}{$0.4$}  &  0.7  &  2.531  &  10.215  &  2.009  &  6.031  &  2.381  &  10.695  &  2.062  &  6.952  \\
  &  0.8  &  2.439  &  10.440  &  2.045  &  7.278  &  2.257  &  11.003  &  1.117  &  8.182  \\
  &  0.9  &  2.336  &  10.659  &  2.088  &  8.317  &  2.130  &  11.274  &  1.173  &  9.063  \\
  &  1.0  &  2.236  &  10.853  &  2.134  &  9.015  &  2.035  &  11.480  &  2.227  &  9.443
\end{tabular}
\end{ruledtabular}
\end{minipage}
\end{table*}

With the specification of the EoS with discontinuous energy density, Eq.~\eqref{EoSEq}, and model parameters $\{\rho_{\rm dis}^+, \alpha, A\}$, the TOV equations given in Eqs.~\eqref{TOV1}-\eqref{TOV2} are integrated from the center at $r=0$, with boundary conditions $m(0)= 0$ and $\rho(0)= \rho_c$, to the surface of the star. Here $\rho_c$ denotes the central energy density and will be varied to construct a sequence of equilibrium stellar configurations. Furthermore, since here we specifically focus on stars presenting a dark energy core described by the Chaplygin fluid, as already discussed, we impose values for the central energy density necessarily in the range of $\rho_c>\rho_{\rm dis}^+$. Of course, being a hybrid system, the set of differential equations \eqref{TOV1}-\eqref{DiffEqPhi} has to be solved separately (i.e., for each phase) so that the different variables must be related by appropriate junction conditions on the domain wall located at $r= R_{\rm dis}$. Once the metric variables ($\Phi$ and $\Psi$) and the thermodynamic quantities ($\rho$ and $p$) have been calculated, we solve Eq.~(\ref{yEq}) to determine the Love number $k_2$ and hence the tidal deformability $\Lambda$.

Similarly for the Sturm-Liouville problem governing the radial pulsations within the stellar fluid, we solve Eqs.~(\ref{ROEq1}) and (\ref{ROEq2}) separately using the above matching conditions at the interface. As specified in Ref.~\cite{Pretel2024}, here we employ the shooting method to find the appropriate eigenfrequencies given a stellar configuration of central density $\rho_c$. The solution to this problem provides a discrete set of eigenvalues $\omega_0^2< \omega_1^2 < \cdots < \omega_n^2 < \cdots$, where $n$ denotes the number of nodes between the center and the surface of the hybrid star.

\begin{figure*}
\centering
\includegraphics[scale=0.73]{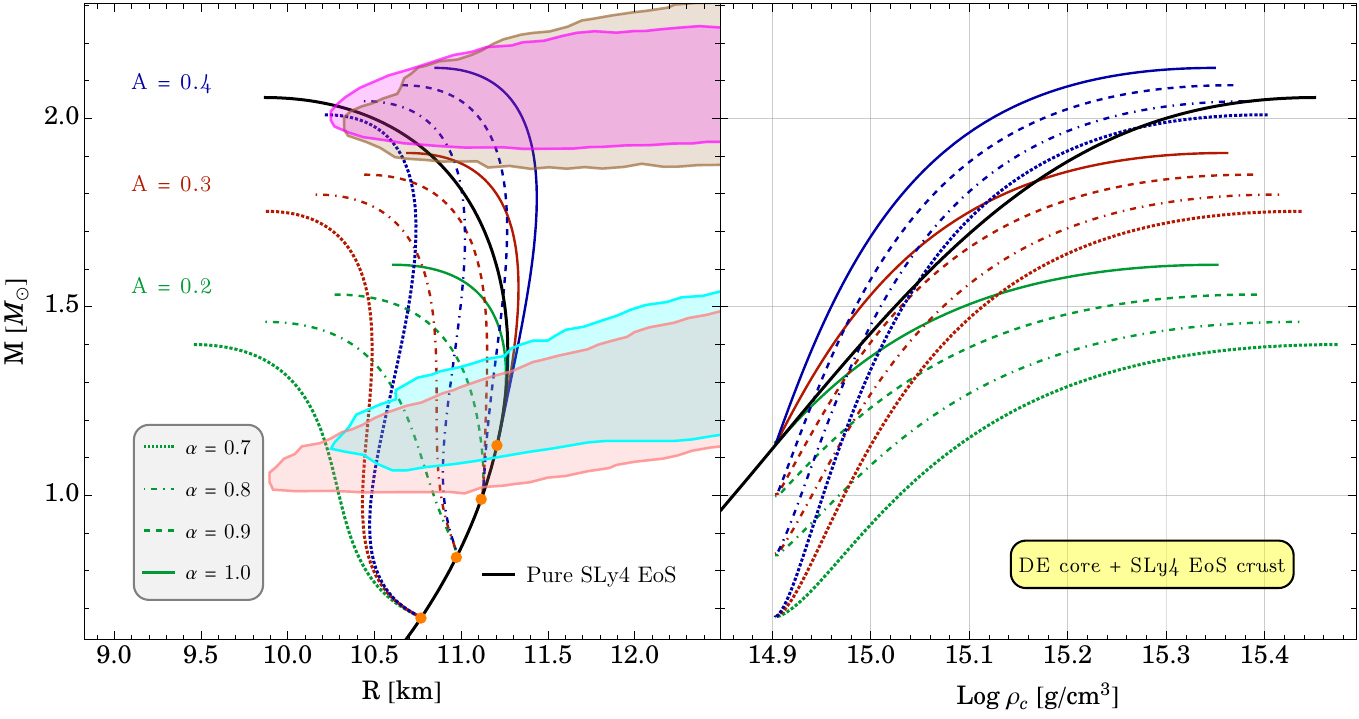}
\caption{Mass-radius diagrams (left), and mass versus central energy density relation (right) for neutron stars with a dark-energy core using inner energy density $\rho_{\rm dis}^+= 0.8 \times 10^{15}\, \rm g/cm^3$. The ordinary-matter crust is described by the SLy4 EoS for four values of the energy density ratio at the discontinuous surface, namely, $\alpha \in [0.7, 1.0]$. The black solid curve has been included as a benchmark for the pure SLy4 EoS and the orange dots on the left plot indicate the transition to the dark sector. Furthermore, for the dimensionless parameter $A$ we have considered three values. The different color contours represent the $90\%$ CI regions obtained from the pulsars PSR J$0030+0451$ \cite{Riley2019, Miller2019} and PSR J$0740+6620$ \cite{Riley2021, Miller2021}. Increasing both $A$ and $\alpha$ leads to higher maximum-mass values. Note also that the blue curves (i.e., when $A=0.4$) consistently satisfy all NICER constraints. Specific values of $\rho_c$ for the maximum mass configurations are given in Table \ref{table}.}
\label{FigMRdSLy4}
\end{figure*}

\begin{figure*}[!htb]
\centering
\includegraphics[scale=0.73]{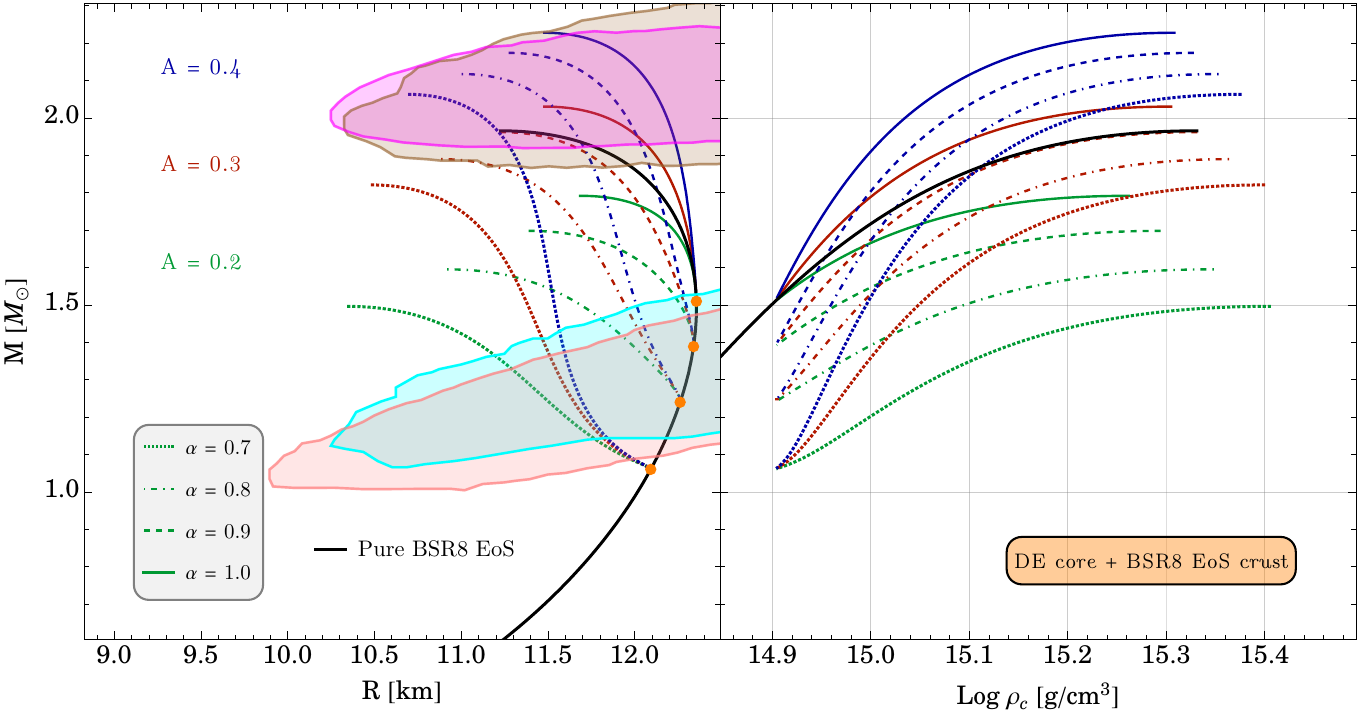}
\caption{Mass as a function of the radius (left) and central density (right) predicted by Eq.~(\ref{EoSEq}) using the BSR8 EoS for the outer layer. Here the black solid curve represents the results obtained using the pure BSR8 EoS. The free parameters $\{\rho_{\rm dis}^+, \alpha, A\}$ and the different color bands are the same as in Fig.~\ref{FigMRdSLy4}.}
\label{FigMRdDEBSR8}
\end{figure*}

The dependence of the total gravitational mass, in solar masses $M_{\odot}$, against the total radius and the central energy density, are respectively shown on the left and right panels of Figs.~\ref{FigMRdSLy4} and \ref{FigMRdDEBSR8}, for an inner energy density $\rho_{\rm dis}^+= 0.8 \times 10^{15}\, \rm g/cm^3$ and different values of~$A$ and~$\alpha$. We assume this value for $\rho_{\rm dis}^+$ in order to be conservative with previous research \cite{Arbanil2023, Pretel2024}, although we must point out that a smaller $\rho_{\rm dis}^+$ may increase the maximum mass of a NS with a CDF core \cite{Pretel2024}. For comparison reasons, we have also included the results associated with single-phase stars (i.e., for normal matter) by black lines in the $M-R$ diagrams, where the orange dots on each left plot indicate the transition to the dark sector.

In Fig.~\ref{FigMRdSLy4}, hybrid stars with dark energy in the core and SLy$4$ EoS on the crust are considered, while in Fig.~\ref{FigMRdDEBSR8} compact stars with a CDF core and BSR$8$ EoS on the crust are taken into account. These $M-R$ results are contrasted with the observational data corresponding to NICER constraint obtained from the pulsars PSR J$0030+0451$ \cite{Riley2019, Miller2019} and PSR J$0740+6620$ \cite{Riley2021, Miller2021}. In the left panel of Figs.~\ref{FigMRdSLy4} and \ref{FigMRdDEBSR8}, the $M(R)$ curves grow from right to left until reaching the maximum mass point and, from this point, the curves turn counterclockwise so that the mass begins to decrease with the total radius of the star. In the right panel of Figs.~\ref{FigMRdSLy4} and \ref{FigMRdDEBSR8}, the curve $M(\rho)$ increases from left to right until reaching the maximum mass value, from this point the mass decreases with the increase of the central energy density. As we will see later in our analysis of radial stability, these points of maximum mass correspond to critical central densities where the stars cease to be stable. The specific values for these critical densities are reported in Table \ref{table}, which decrease as the interface energy density ratio increases in the range $\alpha \in [0.7, 1.0]$ for a fixed $A$.

\begin{figure*}[!htb]
\centering
\includegraphics[scale=0.73]{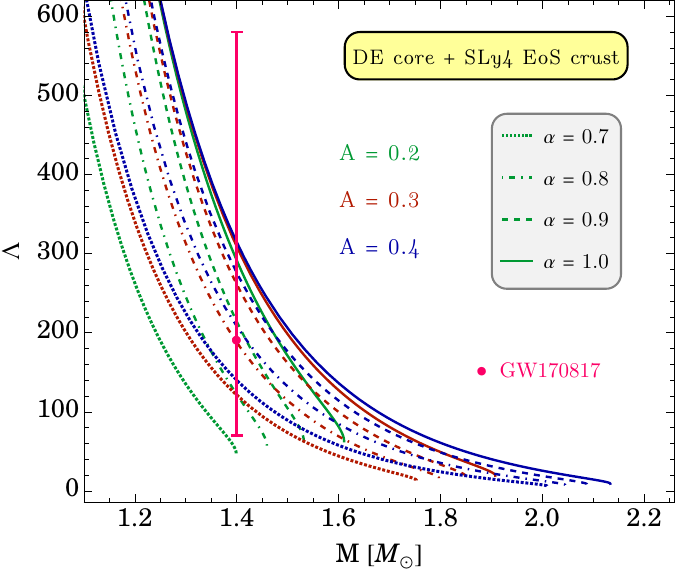}
\includegraphics[scale=0.73]{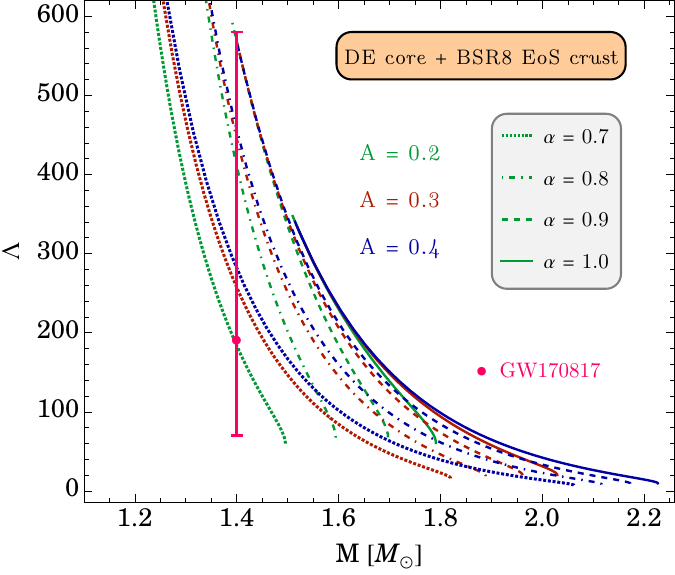}
\caption{The dimensionless tidal deformability $\Lambda$ as a function of the NS mass using the same values for the parameters  $A$ and $\alpha$ as in Figs.~\ref{FigMRdSLy4} and \ref{FigMRdDEBSR8}. The magenta vertical line stands the tidal deformability constraint from the first direct detection of gravitational waves from the coalescence of a NS binary system, i.e.~$\Lambda_{1.4}= 190_{-120}^{+390}$ for the GW170817 signal \cite{Abbott2018PRL}.}
\label{FigLambMasses}
\end{figure*}

From the panel on the left-hand side of Figs.~\ref{FigMRdSLy4} and \ref{FigMRdDEBSR8}, we observe that stars with high masses and their respective radii are sensitive to the change of the parameters $A$ and $\alpha$. For instance, in Fig.~\ref{FigMRdSLy4}, analyzing the maximum total mass value for $\alpha=0.7$, we observe that when $A=0.2$ is incremented in $50\%$ ($A=0.3$) and $100\%$ ($A=0.4$), the total mass grows in approximately $25\%$ and $46\%$, respectively, and their respective radii are increased in $4\%$ and $8\%$. The growth of the total mass and radius with $A$ is understood once observing that larger values of $A$ bring larger total pressure values, see Eq.~\eqref{EoSEq}. In this way, larger pressure can support a more massive star. In addition, examining the maximum total mass for fixed $A=0.2$, we note that when the factor $\alpha =0.7$ grows by $0.1$, $0.2$, and $0.3$, the value of the maximum mass and its respective radius grow in about respectively by $4\%$ and $2\%$, $8\%$ and $11\%$, and $12\%$ and $16\%$. On the other hand, from the panel on the right-hand side of Fig.~\ref{FigMRdSLy4}, when $A$ and $\alpha$ are increased, we see that the maximum total mass is reached in lower central energy density. See also the data in Table~\ref{table} for further comparison details.

The effects of the change of an EoS on the crust can also be seen in Figs.~\ref{FigMRdSLy4} and \ref{FigMRdDEBSR8}. We note that, in our case, when a stiffer parametrization (see also Fig.~\ref{FigEoSs}) is used for the envelope, compact stars with larger total maximum mass and radius are found at lower central energy densities. Such statement is also based on the analysis of two more hadronic models, namely, the SD1 nonrelativistic Skyrme parametrization~\cite{brett-jerome}, and the FSUGZ06 relativistic one~\cite{kumar2006} (results not shown). Particularly, for the some interval of central energy density, we notice that the change of $A$, $\alpha$, and the EoS on the outer layer of the star, allows us to have results closer to the observational evidence of millisecond pulsars PSR J$0030+0451$ and PSR J$0740+6620$. Therefore, when the crust is described by a stiffer hadronic parametrization than the SLy4 one, the $M-R$ curves are shifted to the upper right side generating greater compatibility with observational measurements.

The dimensionless tidal deformability as a function of the total mass is plotted in Fig.~\ref{FigLambMasses} for some values of $A$, $\alpha$, and for the two different EoSs for the NS crust. 

The SLy$4$ and BSR$8$ EoSs for the star's envelope are employed on the left and right-hand panels, respectively. For the range of masses exhibited, we notice that the value of $\Lambda$ grows with the value of the parameters~$A$ and~$\alpha$. Additionally, by changing the EoS of the envelope to a stiffer one, we notice that the value of the deformability grows more noticeably. In the figure, it is also shown the tidal deformability limited by the event GW$170817$ for a $1.4M_{\odot}$ star to be $70\leq\Lambda_{1.4}\leq580$ \cite{Abbott2018PRL}, for low-spin priors, at $90\%$ confidence level. Our $\Lambda-M$ results are consistent with the tidal deformability observed through gravitational-wave emission of coalescing NSs. Nevertheless, by using a stiffer EoS, we find that equilibrium configurations with lower values of $A$ and $\alpha$ can be within the observational data of the GW$170817$ signal. The specific values of $\Lambda$ for the maximum-mass configurations can be found in Table~\ref{table}.

\begin{figure*}[!htb]
\centering
\includegraphics[scale=0.72]{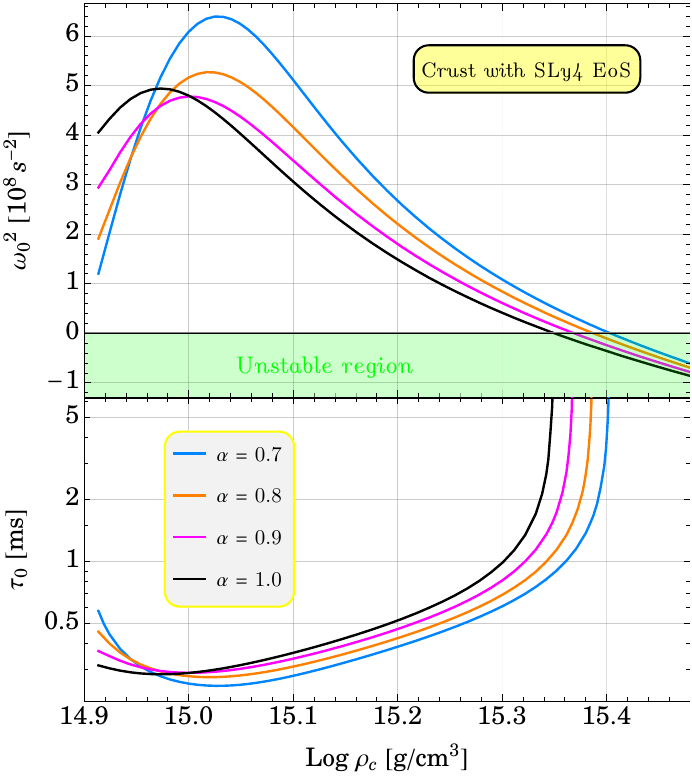}
\includegraphics[scale=0.72]{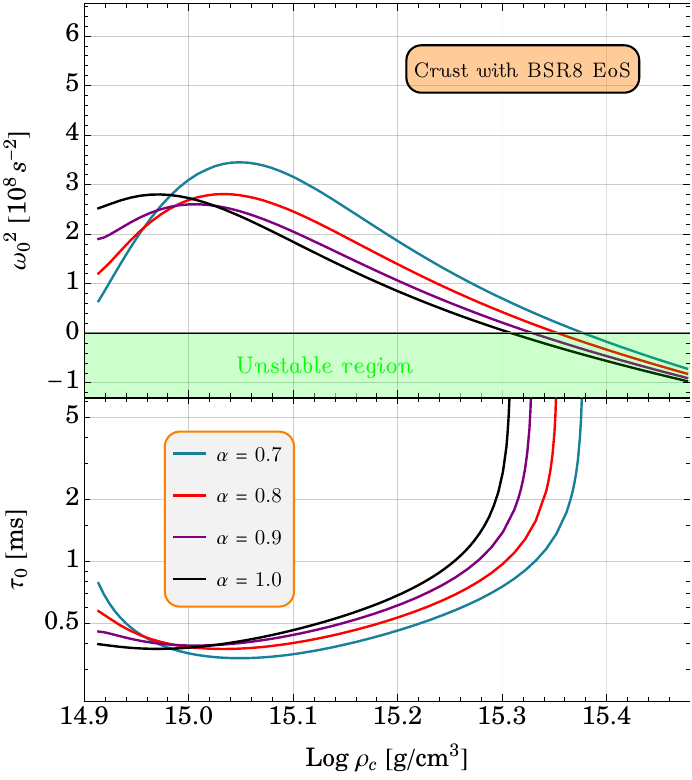}
\caption{Oscillation spectrum under the effect of rapid phase transition for NSs with a dark energy core and SLy4 (left) and BSR8 (right) models for the crust. In each plot, the top panel represents the squared frequency of the fundamental vibration mode $(\omega_0^2)$, while the bottom panel shows the corresponding time taken by the star to complete one oscillation ($\tau_0$), both as functions of the central energy density $\rho_c$. Results obtained with $\rho_{\rm dis}^+= 0.8 \times 10^{15}\, \rm g/cm^3$, $A= 0.4$ and $\alpha \in [0.7, 1.0]$, i.e., blue curves in Figs.~\ref{FigMRdSLy4} and \ref{FigMRdDEBSR8}. Furthermore, the stellar configurations found within the green region are dynamically unstable since $\omega_0^2<0$. }
\label{FigFreqs}
\end{figure*}

In our previous study on NSs with a dark-energy core from the Chaplygin gas~\cite{Pretel2024}, we observed that only rapid phase transitions are compatible with the standard stability criterion $dM/d\rho_c>0$ on the mass versus central density relation, so here we calculate the oscillation spectrum taking into account rapid phase transitions, i.e., by making use of the junction conditions presented in Eq.~\eqref{JuncCond2}. This allows us to rigorously know if the neutron stars admixed with dark energy represented in the $M-R$ diagrams are stable against radial perturbations. Fig.~\ref{FigFreqs} presents the squared frequency (upper) and period (lower) of the fundamental vibration mode as a function of the central density for both EoSs describing the crust. 

\begin{figure*}[!htb]
\centering
\includegraphics[scale=0.72]{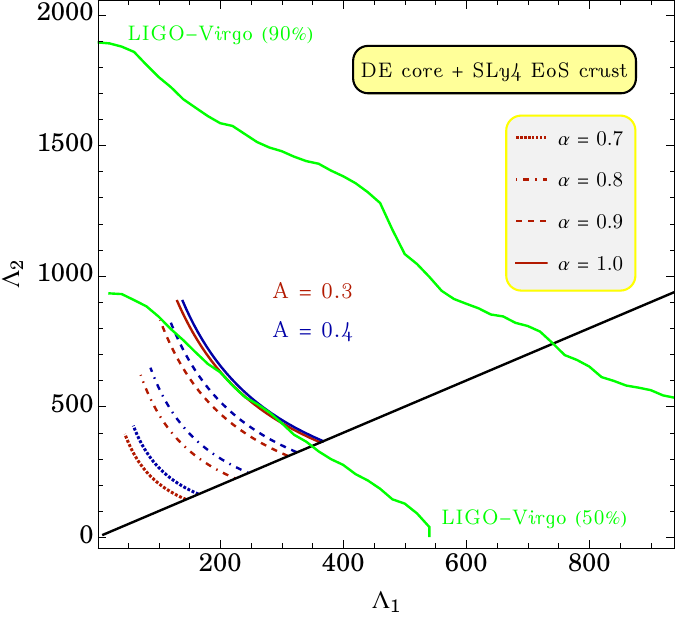}
\includegraphics[scale=0.72]{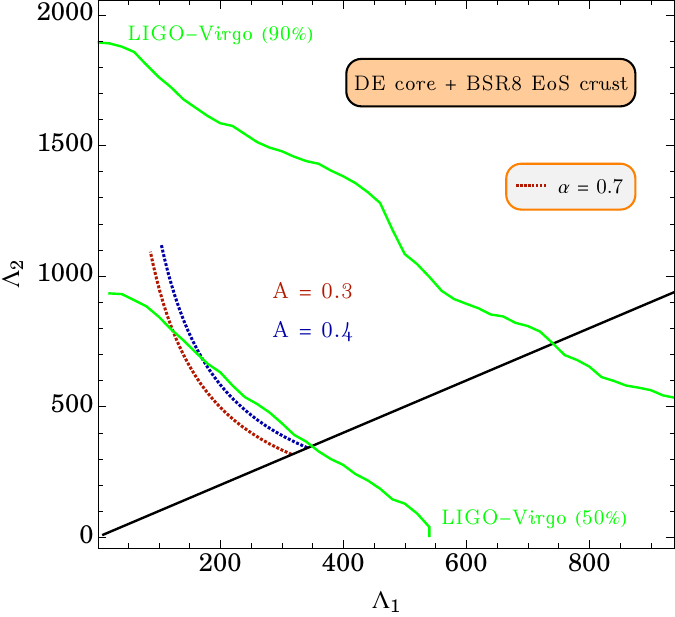}
\caption{Tidal deformabilities of binary NSs with a dark energy core calculated for some values of $A$ and $\alpha$, and considering the SLy4 (left) and BSR8 (right) EoSs for the crust. Note further that our results are compared to the $50\%$ and $90\%$ probability contours (see green lines) given by GW$170817$ event \cite{Abbott2017}. }
\label{FigLamb1Lamb2}
\end{figure*}

Eigenvalues corresponding to $\omega_0^2>0$ indicate radially stable stars, while unstable configurations lie in the negative $\omega_0^2$ region. The period of the fundamental oscillation mode is given by $\tau_0= 2\pi/\omega_0$, and as expected, it tends to infinity as the stars approach the unstable zone (close to the maximum mass configuration). This is the time taken by the star to carry out one radial pulsation and is usually given in milliseconds. Furthermore, we can infer from these results that the larger the energy density jump (i.e., as $\alpha$ decreases) the larger the central density where the change from stability to instability occurs. We can therefore say that, regardless of the EoS for the outer layer, smaller values of $\alpha$ increase the stability of a NS with a CDF core.

The $\Lambda_1\times\Lambda_2$ curves for a binary compact star system with the same chirp mass of the GW$170817$ event are shown in Fig.~\ref{FigLamb1Lamb2} for some values of $A$ and $\alpha$, and the two EoSs for the envelope of the star.

Such curves are found by choosing a value of $m_1$ and, from this choice, we derive $m_2$ for a fixed chirp mass value ${\cal M}=1.188\, M_{\odot}$, which is the most likely value from the parameter estimation reported in \cite{Abbott2017}. This value can be theoretically derived by the equality:
\begin{equation}
{\cal M}=\frac{\left(m_1\,m_2\right)^{3/5}}{\left(m_1+m_2\right)^{1/5}} .   
\end{equation}
The interval considered for the high-mass component $m_1$ and the low-mass component $m_2$ are respectively $1.36\leq m_1/M_{\odot}\leq1.60$ and $1.17\leq m_2/M_{\odot}\leq1.36$. In Fig.~\ref{FigLamb1Lamb2}, we investigate the influence of the parameters $A$ and $\alpha$ and the use of a stiffer EoS on the properties of the compact star. In this situation, it can be seen that these free parameters could play an important role in the detection of these compact objects since their tidal deformabilities are comparable with the $50\%$ and $90\%$ probability contours from GW$170817$ event \cite{Abbott2017}. Through the results shown, one observes that an envelope with SLy4 EoS produces $\Lambda_1\times\Lambda_2$ curves for all values of $\alpha$ adopted in the $M-R$ diagram and are favored by GW$170817$ event. Nonetheless, for the BSR8 EoS crust, only $\alpha= 0.7$ allows us to obtain curves within the detectability range of the LIGO-Virgo Collaboration.

\section{Summary and conclusions}
\label{conclusion}

The most abundant ingredients of the cosmos are dark fluids such as dark matter and dark energy (i.e., those that are invisible but dominate the structure and evolution of the universe), while only a small part is made of ordinary matter. This raises questions about the presence of dark fluids in astrophysical systems such as compact stars. In the absence of a consensus on the theoretical description of the current accelerated expansion of the universe, the Chaplygin gas and its generalized models provide a phenomenologically useful description for modern cosmology, as well as having a well-defined connection with string theory~\cite{Bordemann1993,Ogawa2000,Jackiw2000}. Consequently, a CDF could be a naturally existing substance in the interiors of NSs. Motivated by this, we have studied spherically symmetric systems containing two fluid phases, where the core of the star is a CDF background and the envelope is made of normal matter. The possible existence of these stellar configurations was analyzed through adiabatic radial perturbations and by comparing their macroscopic properties with different observational measurements.

Considering two realistic equations of state for the crust, namely, the parametrizations SLy4 and BSR8 of the Skyrme and RMF models, respectively, we have investigated the effect of a CDF core on the main global properties of NSs such as radius, mass, tidal deformability, and radial oscillation spectrum. In summary, our main findings are the following:
\begin{itemize}
  \item[$i)$] Given a fixed value for the rate of energy densities at the interface $(\alpha)$, increasing the CDF parameter $A$ leads to a significant increase in the total mass of the star. In turn, for fixed $A$, the main consequence of $\alpha$ is a decrease in the mass as the energy density jump becomes larger. The behavior for the radius is less trivial as these parameters vary. Moreover, the EoS models adopted to describe the crust allow to construct compact stars that satisfy the NICER measurements. We further observe that the BSR8 parametrization for the envelope generates larger radii and masses than those caused by using the SLy4 model.
  \item[$ii)$] By means of normal oscillation modes, we have investigated the radial stability of the equilibrium configurations represented in the $M-R$ diagram. Our results show that, regardless of the EoS for the outer layer, smaller values of $\alpha$ increase the stability of a NS with a CDF core. The stars cease to be stable at a critical central density value where the squared frequency of the fundamental mode is zero. We have also calculated the vibration period of these stars for the fundamental mode showing that they take a few milliseconds to complete a full pulsation.
  \item[$iii)$] For stable stars, i.e., those found in the region where $dM/d\rho_c>0$, increasing the CDF parameter $A$ increases the dimensionless tidal deformability $\Lambda$ for fixed $\alpha$ and $M$. Similarly, given a mass $M$ and a fixed $A$, the tidal deformability increases with increasing $\alpha$. Remarkably, these results are in good agreement with the tidal deformability constraint from the GW170817 event. Furthermore, through the $\Lambda_1\times\Lambda_2$ curves, our results for a $1.188\, M_\odot$ chirp-mass NS merger are consistent with such signal too. We also observe that a softer EoS for the crust like the SLy4 EoS fits better with the probability contours reported by the LIGO-Virgo Collaboration.

\end{itemize}

In future studies, it would also be interesting to explore the inverted hybrid case where the star consists of an ordinary matter core and a dark energy shell, as well as the scenario of a neutron star mixed with dark energy generated by a model with both components simultaneously. Such configurations could be part of binary star systems, for instance. \\

\section*{Acknowledgements}

\noindent
JMZP acknowledges support from ``Fundação Carlos Chagas Filho de Amparo à Pesquisa do Estado do Rio de Janeiro'' -- FAPERJ, Process SEI-260003/000308/2024. This work was also supported by Conselho Nacional de Desenvolvimento Cient\'ifico e Tecnol\'ogico (CNPq) under Grants No. 307255/2023-9~(OL), and No. 308528/2021-2~(MD). OL and MD also acknowledge CNPq under Grant No. 401565/2023-8~(Universal). SBD thanks CNPq for partial financial support. JDVA thanks Universidad Privada del Norte and Universidad Nacional Mayor de San Marcos for the financial support - RR Nº$\,005753$-$2021$-R$/$UNMSM under the project number B$21131781$. This work has been done as a part of the Project INCT-Física Nuclear e Aplicações, Project number 464898/2014-5.

\bibliographystyle{apsrev4-2}

\end{document}